\documentclass[twocolumn,aps,prc,showpacs,showkeys,superscriptaddress,nofootinbib]{revtex4}
\usepackage{graphicx}
\usepackage{amsmath}
\usepackage{amsfonts}
\usepackage{amssymb}
\newcommand{\lea}{\raisebox{-.3ex}{\small $ \
\stackrel{\textstyle <}{\sim} $ }}
\newcommand{\gea}{\raisebox{-.3ex}{\small $ \
\stackrel{\textstyle >}{\sim} $ }}
\newcommand{\boldtau}{\mbox{\boldmath $\tau$}}
\newcommand{\beq}{\begin{equation}}
\newcommand{\eeq}{\end{equation}}
\newcommand{\beqa}{\begin{eqnarray}}
\newcommand{\eeqa}{\end{eqnarray}}
\newcommand{\eq}[1]{Eq.~(\ref{#1})}

\date{\today}
\begin{document}

\title{Nonperturbative renormalization of the chiral nucleon-nucleon interaction up to next-to-next-to-leading order}

\author{E. Marji}
\affiliation{Department of Physics, University of Idaho, Moscow, Idaho 83844, USA}
\affiliation{College of Western Idaho, Nampa, Idaho 83653, USA}
\author{A. Canul}
\affiliation{Department of Physics, University of Idaho, Moscow, Idaho 83844, USA}
\author{Q. MacPherson}
\affiliation{Department of Physics, University of Idaho, Moscow, Idaho 83844, USA}
\author{R. Winzer}
\affiliation{Department of Physics, University of Idaho, Moscow, Idaho 83844, USA}
\author{Ch. Zeoli}
\affiliation{Department of Physics, University of Idaho, Moscow, Idaho 83844, USA}
\affiliation{Department of Physics, Florida State University, Tallahassee, Florida 32306, USA}
\author{D. R. Entem}
\email{entem@usal.es}
\affiliation{Grupo de F\'isica Nuclear, IUFFyM, Universidad de Salamanca, E-37008 Salamanca, Spain}
\author{R. Machleidt}
\email{machleid@uidaho.edu}
\affiliation{Department of Physics, University of Idaho, Moscow, Idaho 83844, USA}

\begin{abstract}
We study the
nonperturbative renormalization of the nucleon-nucleon ($NN$) interaction at next-to-leading order (NLO)
and next-to-next-to-leading order (NNLO) of chiral effective field theory.
A systematic variation of the cutoff parameter is performed for values 
below the chiral symmetry breaking scale of about 1 GeV. 
The accuracy of the predictions is determined by calculating the $\chi^2$ for the reproduction 
of the $NN$ data
for energy intervals below pion-production threshold.
At NLO, $NN$ data are described well up to about 100 MeV laboratory energy and,
at NNLO, up to about 200 MeV---with, essentially,
cutoff independence for cutoffs between about 450 and 850 MeV.
\end{abstract}

\maketitle

\section{Introduction}

During the past three decades, it has been demonstrated that chiral effective field theory (chiral EFT) represents 
a powerful tool to deal with hadronic interactions at low energy in a systematic and model-independent
way (see Refs.~\cite{ME11,EHM09} for recent reviews). 
The systematics is provided by
a low-energy expansion arranged in terms of the soft scale over the hard scale,
$(Q/\Lambda_\chi)^\nu$,  where $Q$ is generic for an external
momentum (nucleon three-momentum or pion four-momentum) or a pion mass
and $\Lambda_\chi \approx 1$ GeV the chiral symmetry breaking scale.

The early applications of chiral perturbation theory (ChPT) focused on systems like $\pi\pi$~\cite{GL84}
and $\pi N$~\cite{GSS88}, where the Goldstone-boson character of the pion
guarantees that a perturbative expansion exists.
But the past 20 years have also seen great progress in applying ChPT to nuclear forces
\cite{ME11,EHM09,Wei90,ORK94,KBW97,EGM98,EGM00,EM02a,EM03,EGM05}. 
The nucleon-nucleon ($NN$) system is characterized by large scattering lengths and bound states indicating the nonperturbative
character of the problem. Weinberg~\cite{Wei90} therefore suggested to calculate the nuclear amplitude
in two steps. In step one, the {\it nuclear potential}, $\widehat{V}$, is defined as the sum of irreducible
diagrams, which are evaluated {\it perturbatively} up to the given order. Then in step two, this potential
is iterated to all order (i.e., summed up {\it nonperturbatively}) in the Schr\"odinger or 
Lippmann-Schwinger (LS) equation:
\begin{eqnarray}
 \widehat{T}({\vec p}~',{\vec p}) & = & \widehat{V}({\vec p}~',{\vec p})+
\int d^3p''\:
\widehat{V}({\vec p}~',{\vec p}~'')\:
\nonumber \\ && \times
\frac{M_N}
{{ p}^{2}-{p''}^{2}+i\epsilon}\:
\widehat{T}({\vec p}~'',{\vec p}) \, ,
\label{eq_LS}
\end{eqnarray}
where $\widehat{T}$ denotes the $NN$ T-matrix and $M_N$ the nucleon mass.

In general, the integral in
the LS equation is divergent and needs to be regularized.
Therefore, the potential $\widehat V$
is multiplied
with the regulator function $f(p',p)$,
\begin{equation}
{\widehat V}(\vec{ p}~',{\vec p})
\longmapsto
{\widehat V}(\vec{ p}~',{\vec p}) \, f(p',p) 
\end{equation}
with
\begin{equation}
f(p',p) = \exp[-(p'/\Lambda)^{2n}-(p/\Lambda)^{2n}] \,.
\label{eq_f}
\end{equation}
Typical choices for the cutoff parameter $\Lambda$ that
appears in the regulator are 
$\Lambda \approx 0.5 \mbox{ GeV} < \Lambda_\chi \approx 1$ GeV.

It is pretty obvious that results for the T-matrix may
depend sensitively on the regulator and its cutoff parameter.
This is acceptable if one wishes to build models.
For example, the meson models of the past~\cite{MHE87}
always depended sensitively on the choices for the
cutoffs which, in fact,
were used as additional fit parameters (besides the coupling constants).
However, the EFT approach is expected to be model-independent.

In field theories, divergent integrals are not uncommon and methods have
been developed for how to deal with them.
One regulates the integrals and then removes the dependence
on the regularization parameters (scales, cutoffs)
by renormalization. In the end, the theory and its
predictions do not depend on cutoffs
or renormalization scales.
So-called renormalizable quantum field theories, like QED,
have essentially one set of prescriptions 
that takes care of renormalization through all orders. 
In contrast, 
EFTs are renormalized order by order. 

Weinberg's implicit assumption~\cite{Wei90,Wei09} was that the counterterms
introduced to renormalize the perturbatively calculated
potential, based upon naive dimensional analysis (``Weinberg counting''),
are also sufficient to renormalize the nonperturbative
resummation of the potential in the LS equation.
In 1996, Kaplan, Savage, and Wise (KSW)~\cite{KSW96}
pointed out that there are problems with the Weinberg scheme
if the LS equation is renormalized 
by minimally-subtracted dimensional regularization.
This criticism resulted in a flurry of publications on
the renormalization of the nonperturbative
$NN$ problem.
The literature is too comprehensive
to elaborate on all contributions. Therefore, we will restrict ourselves, here, to discussing
just a few aspects that we perceive as particularly important. 
A more comprehensive consideration can be found in Ref.~\cite{ME11}

Naively, the ideal renormalization procedure is the one where the cutoff
parameter $\Lambda$ is carried to infinity while stable results are maintained.
This was done successfully at leading order (LO) in the work by Nogga {\it et al}~\cite{NTK05}.
At next-to-next-to-leading order (NNLO), the infinite-cutoff renormalization procedure has been investigated 
in~\cite{YEP09} for partial waves with total angular momentum $J\leq 1$ and
in~\cite{VA07} for all partial waves with $J\leq 5$. At next-to-next-to-next-to-leading order (N$^3$LO), 
the $^1S_0$ 
state was considered in Ref.~\cite{Ent08}, and all states up to $J=6$ were investigated in 
Ref.~\cite{ZME12}.
From all of these works, it is evident that no counter term is effective in partial-waves with
short-range repulsion and only a single counter term can effectively be used in
partial-waves with short-range attraction. Thus, for the $\Lambda \rightarrow \infty$
renormalization prescription, even at N$^3$LO, there exists either one or no counter term
per partial-wave state. This is inconsistent with any reasonable power-counting scheme
and prevents an order-by-order improvement of the predictions.

To summarize:
In the infinite-cutoff renormalization scheme, the potential is applied up to unlimited momenta.  However, the EFT this potential is derived from has validity only for momenta smaller than the chiral symmetry breaking scale $\Lambda_{\chi}\approx 1$ GeV.  The lack of order-by-order convergence and discrepancies in lower partial-waves demonstrate that the potential should not be used beyond the limits of the effective theory~\cite{ZME12} (see Ref.~\cite{EG09}
for a related discussion).  The conclusion then is that cutoffs should be limited to $\Lambda\lesssim\Lambda_{\chi}$
(but see also Ref.~\cite{EG12}).

A possible solution of this problem was proposed already in~\cite{NTK05}
and reiterated in a paper by Long and van Kolck~\cite{LK08}.
A calculation of the proposed kind has been performed by
Valderrama~\cite{Val11}, for the $S$, $P$, and $D$ waves. 
The author renormalizes the LO interaction nonperturbatively
and then uses the LO
distorted wave to calculate the two-pion-exchange contributions at NLO and NNLO
perturbatively. It turns out that perturbative renormalizability requires the introduction
of about twice as many counter terms as compared to Weinberg counting, 
which reduces the predictive power. The order-by-order convergence of 
the $NN$ phase shifts appears to be reasonable, but
the cutoffs used in this perturbative summations are rather soft.

However, even if one considers the above method as successful for $NN$ scattering,
there is doubt if the interaction generated in this approach is of any use
for applications in nuclear few- and many-body problems.
In these applications, one would first have to solve the many-body problem
with the re-summed LO interaction, and then add higher order corrections in
perturbation theory.
It was shown in a recent paper~\cite{Mac09} that the renormalized LO
interaction
is characterized by a very large tensor force from one-pion-exchange (1PE). This is no surprise since
LO is renormalized with $\Lambda \rightarrow \infty$ implying that the 1PE,
particulary its tensor force, is totally uncut.
As a consequence of this, the wound integral in nuclear matter, $\kappa$,
comes out to be about 40\%. The hole-line and coupled cluster expansions
are known to converge $\propto \kappa^{n-1}$
with $n$ the number of hole-lines or particles per cluster.
For conventional nuclear forces, the wound integral is typically between 5 and 10\%
and the inclusion of three-body clusters (or three hole-lines) are needed to
obtain converged results in the many-body system.
Thus, if the wound integral is 40\%, probably, up to six hole-lines need to be
included for approximate convergence. Such calculations are not feasible even with
the most powerful computers of today and will not be feasible any time soon.
Therefore, even if the renormalization procedure proposed in~\cite{LK08} will work
for $NN$ scattering, the interaction produced will be highly impractical (to say
the least) in applications in few- and many-body systems because of convergence problems
with the many-body energy and wave functions.

The various problems with the renormalization procedures discussed above
may have a simple common reason:
An EFT that has validity only for momenta $Q < \Lambda_\chi$ is applied such that
momenta $Q \gg \Lambda_\chi$ are heavily involved (because the regulator cutoff parameter
$\Lambda$ is taken to infinity).
A recent paper by Epelbaum and Gegelia~\cite{EG09} illustrates the point:
The authors construct an exactly solvable toy-model that simulates a pionful EFT 
and yields finite results for
$\Lambda \rightarrow \infty$.  However, as it turns out, these
finite results are incompatible with the underlying EFT, while
for cutoffs in the order of the hard scale consistency is maintained.
In simple terms, the point to realize is this: 
If an EFT calculation produces (accidentally) a finite result for
$\Lambda \rightarrow \infty$, then that
does not automatically imply that this result is also meaningful.

This matter is further elucidated in
the lectures by Lepage of 1997~\cite{Lep97}.
Lepage points out that it makes little sense to take the momentum cutoff beyond
the range of validity of the effective theory. By assumption, our data involves energies
that are too low---wave lengths that are too long---to probe the true structure of
the theory at very short distances. When one goes beyond the hard-scale of the theory,
structures are seen that are almost certainly wrong. Thus, results cannot improve
and, in fact, they may degrade or, in more extreme cases, the theory may become
unstable or untunable. In fact, in the $NN$ case, this is what is happening in 
several partial waves (as reported above). Therefore, Lepage
suggests to take the following three steps when building an effective theory:
\begin{enumerate}
\item
Incorporate the correct long-range behavior: The long-range behavior of the underlying
theory must be known, and it must be built into the effective theory. In the case of
nuclear forces, the long-range theory is, of course, well known and 
given by one- and multi-pion exchanges.
\item
Introduce an ultraviolet cutoff to exclude high-momentum states, or, equivalent, to soften the 
short-distance behavior: The cutoff has two effects: First it excludes high-momentum states,
which are sensitive to the unknown short-distance dynamics; only states that we understand
are retained. Second it makes all interactions regular at $r=0$, thereby avoiding infinities.
\item
Add local correction terms (also known as contact or counter terms) 
to the effective Hamiltonian. These mimic the effects of the 
high-momentum states excluded by the cutoff introduced in the previous step.
In the meson-exchange picture, the short-range nuclear force is described by
heavy meson exchange, like the $\rho(770)$ and $\omega(782)$. However, at low
energy, such structures are not resolved. Since we must include contact terms 
anyhow, it is most efficient to use them to account for any heavy-meson exchange
as well.
The correction terms systematically remove dependence on the cutoff.
\end{enumerate}

Crucial for an EFT are regulator independence (within the range of validity
of the EFT) and a power counting scheme that allows for order-by-order
improvement with decreasing truncation error.
The purpose of renormalization is to achieve this regulator independence while maintaining
a functional power counting scheme.

Thus, in the spirit of Lepage~\cite{Lep97}, the cutoff independence should be examined
for cutoffs below the hard scale and not beyond. Ranges of cutoff independence within the
theoretical error are to be identified.
In this paper, we will present
a systematic investigation of this kind.
In our work, we quantify the error of the predictions by calculating the $\chi^2$/datum 
for the reproduction of the neutron-proton ($np$) elastic scattering data
as a function of the cutoff parameter $\Lambda$ of the regulator
Eq.~(\ref{eq_f}). We will investigate the predictions by chiral $np$ potentials at 
order NLO and NNLO applying Weinberg counting for the counter terms ($NN$ contact terms).

This paper is organized as follows.
In Sec.~II, we present the mathematical formalism which defines chiral $NN$ potentials up to NNLO
and $NN$ scattering. In Sec.~III, we discuss in detail the nonperturbative renormalization
of these potentials and present our results. The paper is concluded in Sec.~IV.

\section{Chiral $NN$ potential and two-nucleon scattering}
\label{sec_NNpot}

Nuclear potentials are defined as sets of irreducible
graphs up to a given order.
The power $\nu$ of a few-nucleon diagram involving $A$ nucleons
is given in terms of naive dimensional analysis by:
\begin{equation} 
\nu = -2 +2A - 2C + 2L 
+ \sum_i \Delta_i \, ,  
\label{eq_nu} 
\end{equation}
with
\begin{equation}
\Delta_i  \equiv   d_i + \frac{n_i}{2} - 2  \, ,
\label{eq_Deltai}
\end{equation}
where $C$ denotes the number of separately connected pieces and
$L$ the number of loops in the diagram;
$d_i$ is the number of derivatives or pion-mass insertions and $n_i$ the number of nucleon fields 
(nucleon legs) involved in vertex $i$; the sum runs over all vertices contained
in the diagram under consideration.
Note that $\Delta_i \geq 0$
for all interactions allowed by chiral symmetry.
For an irreducible $NN$ diagram (``two-nucleon potential'', $A=2$, $C=1$),
Eq.~(\ref{eq_nu}) collapses to
\begin{equation} 
\nu =  2L + \sum_i \Delta_i \, .  
\label{eq_nunn} 
\end{equation}

Thus, in terms of naive dimensional analysis or ``Weinberg counting'',
the various orders of the irreducible graphs which define the chiral $NN$ potential 
are given by:
\beqa
V_{\rm LO} & = & 
V_{\rm ct}^{(0)} + 
V_{1\pi}^{(0)} 
\label{eq_VLO}
\\
V_{\rm NLO} & = & V_{\rm LO} +
V_{\rm ct}^{(2)} + 
V_{1\pi}^{(2)} +
V_{2\pi}^{(2)} 
\label{eq_VNLO}
\\
V_{\rm NNLO} & = & V_{\rm NLO} +
V_{1\pi}^{(3)} + 
V_{2\pi}^{(3)} 
\label{eq_VNNLO}
\eeqa
where 
the superscript denotes the order $\nu$ of the low-momentum
expansion.
Contact potentials carry the subscript ``ct'' and
pion-exchange potentials can be identified by an
obvious subscript.

The charge-independent 1PE potential reads
\begin{equation}
V_{1\pi} ({\vec p}~', \vec p) = - 
\frac{g_A^2}{4f_\pi^2}
\: 
\boldtau_1 \cdot \boldtau_2 
\:
\frac{
\vec \sigma_1 \cdot \vec q \,\, \vec \sigma_2 \cdot \vec q}
{q^2 + m_\pi^2} 
\,,
\label{eq_1PEci}
\end{equation}
where ${\vec p}~'$ and $\vec p$ designate the final and initial
nucleon momenta in the center-of-mass system (CMS) and $\vec q \equiv
{\vec p}~' - \vec p$ is the momentum transfer; $\vec \sigma_{1,2}$ and
$\boldtau_{1,2}$ are the spin and isospin operators of nucleon 1 and
2; $g_A$, $f_\pi$, and $m_\pi$ denote axial-vector coupling constant,
the pion decay constant, and the pion mass, respectively. We use
$f_\pi=92.4$ MeV and $g_A=1.29$ throughout this work.  
Since higher order corrections contribute only to mass
and coupling constant renormalizations and since, on shell, there are
no relativistic corrections, the on-shell 1PE has the form
\eq{eq_1PEci} in all orders.

In this paper, we will specifically calculate neutron-proton ($np$) scattering 
and take the charge-dependence (isospin violation) of the 1PE into account.
Thus, the 1PE potential that we actually apply reads
\begin{equation}
V_{1\pi}^{(np)} ({\vec p}~', \vec p) 
= -V_{1\pi} (m_{\pi^0}) + (-1)^{I+1}\, 2\, V_{1\pi} (m_{\pi^\pm})
\,,
\label{eq_1penp}
\end{equation}
where $I$ denotes the isospin of the two-nucleon system and
\begin{equation}
V_{1\pi} (m_\pi) \equiv - \,
\frac{g_A^2}{4f_\pi^2} \,
\frac{
\vec \sigma_1 \cdot \vec q \,\, \vec \sigma_2 \cdot \vec q}
{q^2 + m_\pi^2} 
\,.
\end{equation}
We use $m_{\pi^0}=134.9766$ MeV and
 $m_{\pi^\pm}=139.5702$ MeV.

\subsection{Leading order (LO)}
The LO chiral $NN$ potential consists of a contact part and an 1PE part,
cf.\ Eq.~(\ref{eq_VLO}). The 1PE part is given by Eq.~(\ref{eq_1penp})
and the LO contacts are
\begin{equation}
V_{\rm ct}^{(0)}(\vec{p'},\vec{p}) =
C_S +
C_T \, \vec{\sigma}_1 \cdot \vec{\sigma}_2 \, ,
\label{eq_ct0}
\end{equation}
and, in terms of partial waves,
\beqa
V_{\rm ct}^{(0)}(^1 S_0)          &=&  \widetilde{C}_{^1 S_0} =
4\pi\, ( C_S - 3 \, C_T )
\nonumber \\
V_{\rm ct}^{(0)}(^3 S_1)          &=&  \widetilde{C}_{^3 S_1} =
4\pi\, ( C_S + C_T ) \,,
\label{eq_ct0_pw}
\eeqa
where $C_S, C_T, \widetilde{C}_{^1 S_0}, \widetilde{C}_{^3 S_1}$ are constants
to be adjusted to $NN$ data.

\subsection{Next-to-leading order (NLO)}

For the NLO chiral $NN$ potential, Eq.~(\ref{eq_VNLO}), we need to specify
the second order contact part and the two-pion exchange (2PE) part.
The NLO contact terms are given by~\cite{ME11}
\beqa
V_{\rm ct}^{(2)}(\vec{p'},\vec{p}) &=&
C_1 \, q^2 +
C_2 \, k^2 
\nonumber 
\\ &+& 
\left(
C_3 \, q^2 +
C_4 \, k^2 
\right) \vec{\sigma}_1 \cdot \vec{\sigma}_2 
\nonumber 
\\
&+& C_5 \left( -i \vec{S} \cdot (\vec{q} \times \vec{k}) \right)
\nonumber 
\\ &+& 
 C_6 \, ( \vec{\sigma}_1 \cdot \vec{q} )\,( \vec{\sigma}_2 \cdot 
\vec{q} )
\nonumber 
\\ &+& 
 C_7 \, ( \vec{\sigma}_1 \cdot \vec{k} )\,( \vec{\sigma}_2 \cdot 
\vec{k} ) \,,
\label{eq_ct2}
\eeqa
with the partial-wave decomposition
\beqa
V_{\rm ct}^{(2)}(^1 S_0)          &=&  C_{^1 S_0} ( p^2 + {p'}^2 ) 
\nonumber \\
V_{\rm ct}^{(2)}(^3 P_0)          &=&  C_{^3 P_0} \, p p'
\nonumber \\
V_{\rm ct}^{(2)}(^1 P_1)          &=&  C_{^1 P_1} \, p p' 
\nonumber \\
V_{\rm ct}^{(2)}(^3 P_1)          &=&  C_{^3 P_1} \, p p' 
\nonumber \\
V_{\rm ct}^{(2)}(^3 S_1)          &=&  C_{^3 S_1} ( p^2 + {p'}^2 ) 
\nonumber \\
V_{\rm ct}^{(2)}(^3 S_1- ^3 D_1)  &=&  C_{^3 S_1- ^3 D_1}  p^2 
\nonumber \\
V_{\rm ct}^{(2)}(^3 D_1- ^3 S_1)  &=&  C_{^3 S_1- ^3 D_1}  {p'}^2 
\nonumber \\
V_{\rm ct}^{(2)}(^3 P_2)          &=&  C_{^3 P_2} \, p p'   \,.
\label{eq_ct2_pw}
\eeqa
To state the 2PE potentials, we introduce the following scheme:
\begin{eqnarray} 
V_{2\pi}^{(\nu)}({\vec p}~', \vec p) &  = &
 \:\, V_C^{(\nu)} \:\, + \,  \bbox{\tau}_1 \cdot \bbox{\tau}_2 
\, W_C^{(\nu)} 
\nonumber \\ &+&  
\left[ \, V_S^{(\nu)}  + \bbox{\tau}_1 \cdot \bbox{\tau}_2 \, W_S^{(\nu)} \right] \,
\vec\sigma_1 \cdot \vec \sigma_2
\nonumber \\ &+& 
\left[ \, V_{LS}^{(\nu)} + \bbox{\tau}_1 \cdot \bbox{\tau}_2 \, W_{LS}^{(\nu)}  \right] \,
\left(-i \vec S \cdot (\vec q \times \vec k) \,\right)
\nonumber \\ &+& 
\left[ \, V_T^{(\nu)}   + \bbox{\tau}_1 \cdot \bbox{\tau}_2 \, W_T^{(\nu)}  \right] \,
\vec \sigma_1 \cdot \vec q \,\, \vec \sigma_2 \cdot \vec q  
\nonumber \\ &+& 
\left[  V_{\sigma L}^{(\nu)} + \bbox{\tau}_1 \cdot \bbox{\tau}_2  
      W_{\sigma L}^{(\nu)}  \right] 
\vec\sigma_1\cdot(\vec q\times \vec k\,) \,
\vec \sigma_2 \cdot(\vec q\times \vec k\,) ,
\nonumber \\ 
\label{eq_nnamp}
\end{eqnarray}
where 
\begin{equation}
\begin{array}{llll}
\vec k &\equiv& \frac12 ({\vec p}~' + \vec p) & \mbox{\rm is the 
average momentum, and}\\
\vec S &\equiv& \frac12 (\vec\sigma_1+\vec\sigma_2) & 
\mbox{\rm the total spin.}
\end{array}
\label{eq_defqk}
\end{equation}
Using the above notation, the NLO 2PE is simply given by~\cite{ME11,KBW97}
\begin{eqnarray} 
W_C^{(2)} &=&-{L(q)\over384\pi^2 f_\pi^4} 
\left[4m_\pi^2(5g_A^4-4g_A^2-1)
\right. \nonumber \\  && \left.
+q^2(23g_A^4 -10g_A^2-1) 
+ {48g_A^4 m_\pi^4 \over w^2} \right] ,  
\label{eq_2C}
\\   
V_T^{(2)} &=& -{1\over q^2} V_{S}^{(2)} 
    \; = \; -{3g_A^4 L(q)\over 64\pi^2 f_\pi^4} \,.
\label{eq_2T}
\end{eqnarray}  
It is well known that the 2PE at NNLO shows an unphysically
strong attraction in the high-momentum components of the loop
integrals, when regularized by dimensional regularization. 
For a realistic 2PE contribution at order NNLO,
it is therefore necessary to cut out those short-range components.
This can be achieved by applying the so-called spectral function
regularization (SFR) in those loop integrals (see Ref.~\cite{EGM04}
for details), which leads to the following
loop function:
\begin{equation} 
L(q)  \equiv \theta(\tilde{\Lambda} -2m_\pi) {w\over 2q} 
\ln {\frac{\tilde{\Lambda}^2w^2+q^2s^2+2\tilde{\Lambda}qws}{4m_\pi^2(\tilde{\Lambda}^2+q^2)}}
\label{eq_L}
\end{equation}
with
\begin{eqnarray} 
 w  & \equiv &  \sqrt{4m_\pi^2+q^2} \;\;\; \mbox{ and} \\
 s & \equiv & \sqrt{\tilde{\Lambda}^2-4m_\pi^2} \,.  
\end{eqnarray}
Note that
\begin{equation}
\lim_{\tilde{\Lambda} \rightarrow \infty} L(q) =  {w\over q} 
\ln {\frac{w+q}{2m_\pi}} \,,
\end {equation}
which is the loop function used in dimensional regularization. 
SFR introduces the cutoff parameter $\tilde{\Lambda}$,
for which a value below the chiral-symmetry breaking scale
of $\sim 1$~GeV is appropriate. For our choices for  $\tilde{\Lambda}$, see below.
SFR is not really necessary for the NLO contribution, but since we
have to use it for NNLO, we apply this also at NLO
for reasons of consistency.

\subsection{Next-to-next-to-leading order (NNLO)}
There are no new contacts at NNLO, cf.\ Eq.~(\ref{eq_VNNLO}), and, thus,
all we need is the third order 2PE potential, which is 
[using the notation introduced in Eq.~(\ref{eq_nnamp})]~\cite{ME11,KBW97}
\begin{eqnarray}
V_C^{(3)} &=& V_{C1}^{(3)} + V_{C2}^{(3)}\,,\\
W_C^{(3)} &=& W_{C1}^{(3)} + W_{C2}^{(3)}\,,\\
V_T^{(3)} &=& V_{T1}^{(3)} + V_{T2}^{(3)}\,,\\
W_T^{(3)} &=& W_{T1}^{(3)} + W_{T2}^{(3)}\,,\\
V_S^{(3)} &=& V_{S1}^{(3)} + V_{S2}^{(3)}\,,\\
W_S^{(3)} &=& W_{S1}^{(3)} + W_{S2}^{(3)}\,,\\
V_{LS}^{(3)} &=&  {3g_A^4  \widetilde{w}^2 A(q) \over 32\pi M_N f_\pi^4} 
 \,,\\  
W_{LS}^{(3)} &=& {g_A^2(1-g_A^2)\over 32\pi M_N f_\pi^4} 
w^2 A(q) \,, 
\end{eqnarray}
where
\begin{eqnarray} 
V_{C1}^{(3)} &=&{3g_A^2 \over 16\pi f_\pi^4} 
\left\{ {g_A^2 m_\pi^5  \over 16M_N w^2}  
\right. \nonumber \\  & - & \left.
\left[2m_\pi^2( 2c_1-c_3)-q^2  \left(c_3 +{3g_A^2\over16M_N}\right)
\right]
\widetilde{w}^2 A(q) \right\} , 
\nonumber \\
\label{eq_3C}
\\
W_{C1}^{(3)} &=& {g_A^2\over128\pi M_N f_\pi^4} \left\{ 
 3g_A^2 m_\pi^5 w^{-2} 
 \right. \nonumber \\  && \left.
 - \left[ 4m_\pi^2 +2q^2-g_A^2(4m_\pi^2+3q^2) \right] 
\widetilde{w}^2 A(q)
\right\} ,
\\ 
V_{T1}^{(3)} &=& -{1 \over q^2} V_{S1}^{(3)}
   \; = \; {9g_A^4 \widetilde{w}^2 A(q) \over 512\pi M_N f_\pi^4} 
 \,,  \\ 
W_{T1}^{(3)} &=&-{1\over q^2}W_{S1}^{(3)} 
    =-{g_A^2 A(q) \over 32\pi f_\pi^4}
     \nonumber \\  & \times & 
\left[
\left( c_4 +{1\over 4M_N} \right) w^2
-{g_A^2 \over 8M_N} (10m_\pi^2+3q^2)  \right] ,
\label{eq_3T}
\end{eqnarray}   
and
\begin{eqnarray}
V_{C2}^{(3)} &=& -\frac{3 g_A^4}{256 \pi f_\pi^4 M_N} 
(m_\pi w^2 + \widetilde w^4 A(q) )
\,,
\label{eq_3EM1}
\\
W_{C2}^{(3)} &=& \frac{g_A^4}{128 \pi f_\pi^4 M_N} 
(m_\pi w^2 + \widetilde w^4 A(q) )
\,,
\\
V_{T2}^{(3)} &=& -\frac{1}{q^2} V_{S2}^{(3)} = \frac{3 g_A^4}{512 \pi f_\pi^4 M_N} 
(m_\pi + w^2 A(q) )
\,,
\\
W_{T2}^{(3)} &=& -\frac{1}{q^2} W_{S2}^{(3)} = -\frac{g_A^4}{256 \pi f_\pi^4 M_N} 
(m_\pi + w^2 A(q) ) \,,
\nonumber \\
\label{eq_3EM4}
\end{eqnarray}
with
\begin{equation} 
\widetilde{w} \equiv  \sqrt{2m_\pi^2+q^2} \,. 
\end{equation}
The loop function that appears in the above expressions,
regularized by SFR, is
\begin{equation} 
A(q) \equiv \theta(\tilde{\Lambda} -2m_\pi) {1\over 2q}
\arctan{q (\tilde{\Lambda}-2m_\pi) \over q^2+2\tilde{\Lambda} m_\pi} \,.
\label{eq_A}
\end{equation}
\begin{equation}
\lim_{\tilde{\Lambda} \rightarrow \infty} A(q) =  
{1\over 2q} \arctan{q \over 2m_\pi} 
\end {equation}
yields the loop function used in dimensional regularization.
Note that Eqs.~(\ref{eq_3EM1})-(\ref{eq_3EM4}) are corrections of the iterative 2PE, see Ref.~\cite{ME11} for details.
In all 2PE potentials, we apply the average nucleon mass, $M_N=938.9182$ MeV, and the
average pion mass, $m_\pi=138.039$ MeV.
For the dimension-two low-energy constants we use 
$c_1=-0.81$ GeV$^{-1}$,
$c_3=-3.40$ GeV$^{-1}$, and
$c_4=3.40$ GeV$^{-1}$~\cite{BM00}.

\subsection{$NN$ scattering}
\label{sec_NN}

For the unitarizing scattering equation, we choose the relativistic three-dimensional equation
proposed by Blankenbecler and Sugar (BbS)~\cite{BS66},
which reads,
\begin{eqnarray}
{T}({\vec p}~',{\vec p}) & = &  {V}({\vec p}~',{\vec p})+
\int \frac{d^3p''}{(2\pi)^3} \:
{V}({\vec p}~',{\vec p}~'') \:
\nonumber \\ && \times
\frac{M_N^2}{E_{p''}} \:  
\frac{1}{{ p}^{2}-{p''}^{2}+i\epsilon} \:
{T}({\vec p}~'',{\vec p}) 
\label{eq_bbs2}
\end{eqnarray}
with $E_{p''}\equiv \sqrt{M_N^2 + {p''}^2}$.
The advantage of using a relativistic scattering equation is that it automatically
includes relativistic corrections to all orders. Thus, in the scattering equation,
no propagator modifications are necessary when raising the order to which the
calculation is conducted.

Defining
\begin{equation}
\widehat{V}({\vec p}~',{\vec p})
\equiv 
\frac{1}{(2\pi)^3}
\sqrt{\frac{M_N}{E_{p'}}}\:  
{V}({\vec p}~',{\vec p})\:
 \sqrt{\frac{M_N}{E_{p}}}
\label{eq_minrel1}
\end{equation}
and
\begin{equation}
\widehat{T}({\vec p}~',{\vec p})
\equiv 
\frac{1}{(2\pi)^3}
\sqrt{\frac{M_N}{E_{p'}}}\:  
{T}({\vec p}~',{\vec p})\:
 \sqrt{\frac{M_N}{E_{p}}}
\,,
\label{eq_minrel2}
\end{equation}
where the factor $1/(2\pi)^3$ is added for convenience,
the BbS equation collapses into the usual, nonrelativistic
Lippmann-Schwinger (LS) equation, Eq.~(\ref{eq_LS}).
Since $\widehat V$ satisfies Eq.~(\ref{eq_LS}), 
it can be used like a usual nonrelativistic potential, and 
$\widehat{T}$ may be perceived as the conventional nonrelativistic T-matrix.
The square-root factors in Eqs.~(\ref{eq_minrel1}) and (\ref{eq_minrel2})
are applied to the potentials of all orders.

In the LS equation, \eq{eq_LS}, we use
\beqa
M_N  &=&  \frac{2M_pM_n}{M_p+M_n} = 938.9182 \mbox{ MeV, and}
\\
p^2 & = & \frac{M_p^2 T_{\rm lab} (T_{\rm lab} + 2M_n)}
               {(M_p + M_n)^2 + 2T_{\rm lab} M_p}  
\,,
\eeqa
where $M_p=938.2720$ MeV and $M_n=939.5653$ MeV
are the proton and neutron masses, respectively, and 
$T_{\rm lab}$ 
is the kinetic energy of the incident neutron 
in the laboratory system (``Lab.\ Energy'').
The relationship between $p^2$ and
$T_{\rm lab}$ 
is based upon relativistic kinematics.

\begin{figure*}
\vspace*{-1cm}
\hspace*{-1.5cm}
\scalebox{0.5}{\includegraphics{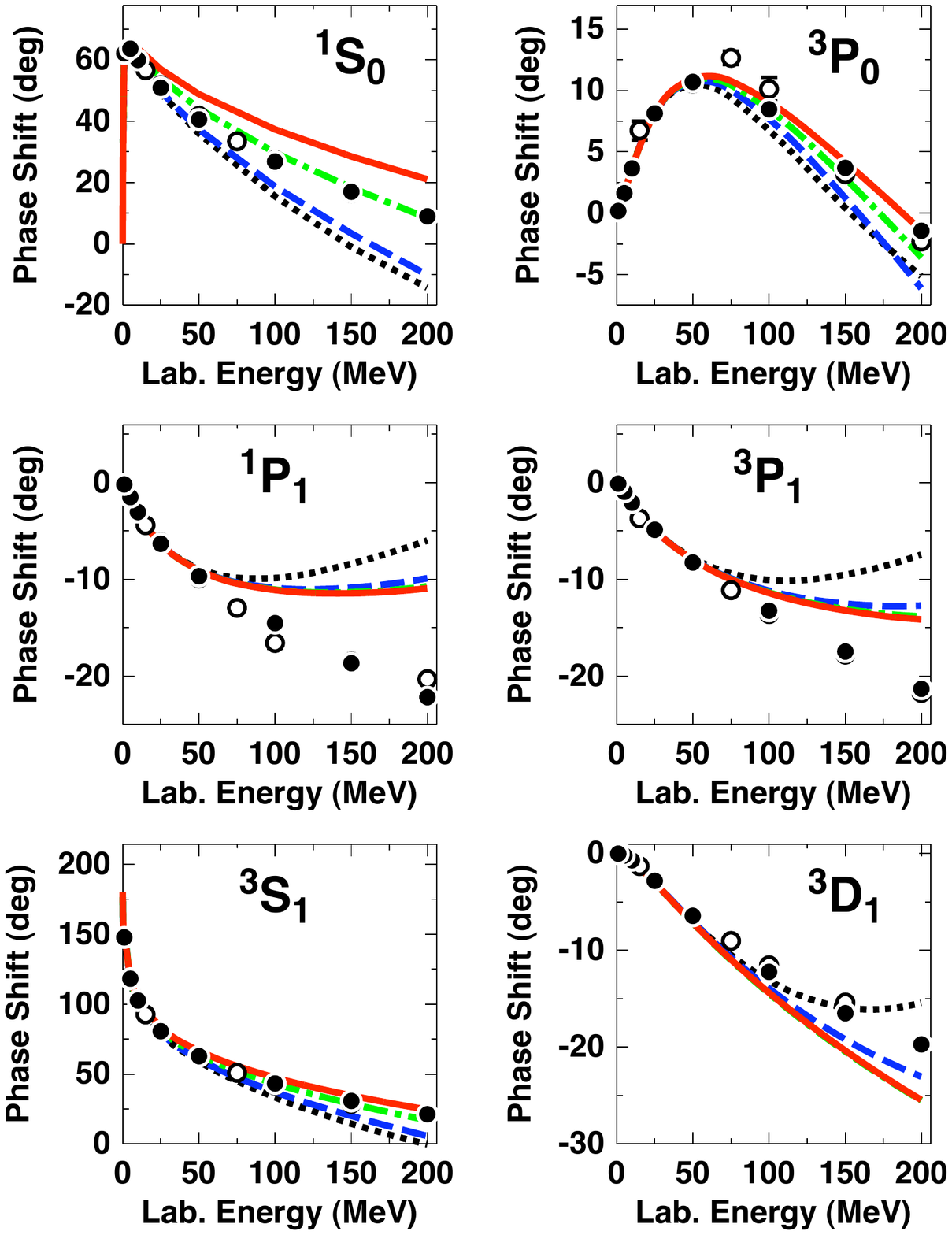}}
\hspace*{-2.5cm}
\scalebox{0.5}{\includegraphics{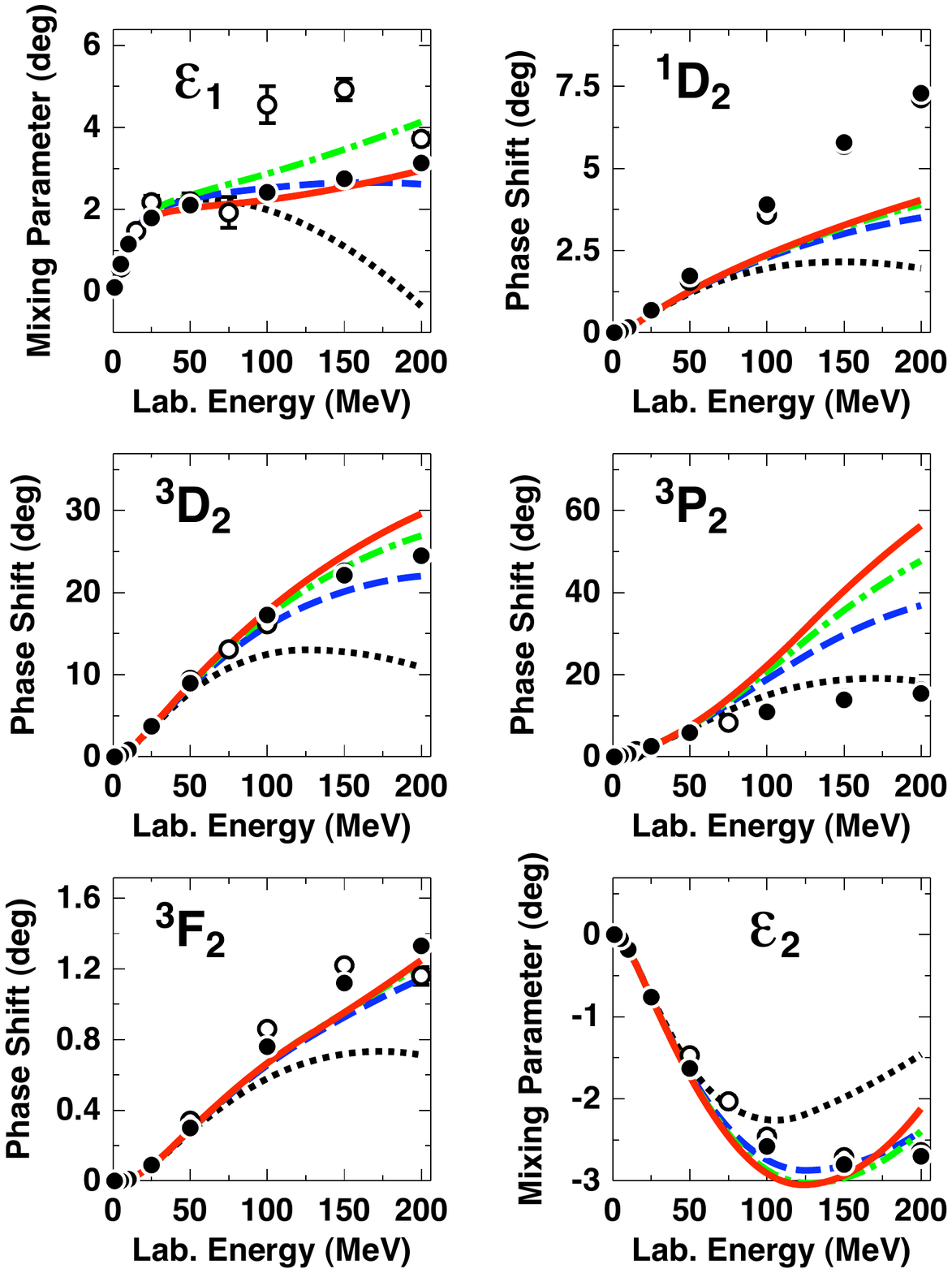}}
\vspace*{-2.5cm}
\caption{(Color online).
Phase-shifts of neutron-proton scattering at order NLO for total angular momentum $J\leq2$ and laboratory kinetic energies below $200\text{\ MeV}$.  The black dotted curve is obtained for $\Lambda=400\text{\ MeV}$, the blue dashed curve for $\Lambda=600\text{\ MeV}$, the green dash-dot curve for $\Lambda=800\text{\ MeV}$, and the red solid curve for $\Lambda=1000\text{\ MeV}$. 
The value for the SFR parameter is $\tilde{\Lambda} = 700$ MeV.
The filled and open circles represent the results from the Nijmegan multi-energy $np$ phase-shift analysis~\cite{Sto93} and the VPI/GWU single-energy $np$ analysis SM99~\cite{SM99}, respectively.
\label{fig_ph}}
\end{figure*}

\section{Nonperturbative renormalization}
\label{sec_reno}

The chiral $NN$ potential depends on two cutoff parameters, $\tilde{\Lambda}$ and ${\Lambda}$.
The parameter $\tilde{\Lambda}$ is introduced for the SFR
of the pion loop integrals. The SFR cuts out the short range part of the 2PE.
To be more specific: Short distance contributions, which are equivalent to contributions
from ``meson masses'' $m>\tilde{\Lambda}$, are set to zero. 
This is necessary, because the short-range part of the 2PE
shows unphysically strong attraction, particularly, at NNLO. 
All this is part of the evaluation of the potential which is calculated and renormalized perturbatively.
It has nothing to do with nonperturbative regularization and renormalization.
The parameter $\tilde{\Lambda}$ is given a fixed value in the order of the $\rho$ mass
or slightly below.

The process of interest in this paper is the
re-summation of the potential in the LS equation (\ref{eq_LS}),
which is a nonperturbative process.
To avoid divergences in the re-summation, the regulator Eq.~(\ref{eq_f}) is applied,
which is a function of the cutoff parameter $\Lambda$. 
The renormalization of this regularized LS equation is the focus of the present investigation.
Successful renormalization means independence of the predictions  from $\Lambda$ 
within the accuracy of the given order.
To investigate this issue, we  vary $\Lambda$ from about 350 MeV to 900 MeV.

We choose this range of cutoff values for the following reasons.
Choosing $\Lambda$ too small removes the long-distance physics in which we trust. 
We certainly want to preserve
the chiral 2PE contribution to the $NN$ interaction. Two pions have a rest-mass of about 280 MeV. 
In addition, the pions have kinetic energy. This suggests 350 MeV as a reasonable guess for a
lower limit, but see discussion below.
The upper limit for cutoffs is dictated by the fact that $\Lambda$ should be below the
chiral symmetry breaking scale $\Lambda_\chi \sim 1$ GeV.

For each value of $\Lambda$, the $S$- and $P$-waves are readjusted with the help
of the seven counter terms available at NLO and NNLO,  cf.\ Eqs.~(\ref{eq_ct0_pw}) and (\ref{eq_ct2_pw}).
$S$-waves carry two counter terms each.
In $^1S_0$, we use them to adjust the scattering length, $a_s$, and the effective range, $r_s$,
to their empirical values,
\begin{eqnarray}
a_s & = & -23.74 \mbox{ fm} \,, \\
r_s & = & 2.70 \mbox{ fm} \,.
\end{eqnarray}
While it is always possible to fit $a_s$ accurately, there are problems fitting $r_s$ for the
higher values of $\Lambda$, in which case we fit $r_s$ as close as possible.

In $^3S_1$, we fit the deuteron binding energy, $B_d$, and the effective range, $r_t$,
\begin{eqnarray}
B_d & = & 2.224575 \mbox{ MeV} \,, \\
r_t & = & 1.75 \mbox{ fm} \,.
\end{eqnarray}
Similar to the $^1S_0$ case, the $r_t$ cannot be fit accurately for the larger values of $\Lambda$,
where we then fit $r_t$ as close as possible. $B_d$ is always reproduced accurately.

$P$-waves have one counter term each, which we utilize to fit 
the empirical phase shifts at 25 MeV as determined in
the Nijmegen multienergy $np$ phase shift analysis~\cite{Sto93}, which are:
\begin{eqnarray}
\delta_{^1 P_1}  (25\mbox{ MeV}) &=& -6.31 \mbox{ deg,}
\\
\delta_{^3 P_0} (25\mbox{ MeV}) &=& 8.13 \mbox{ deg,}
\\
\delta_{^3 P_1}  (25\mbox{ MeV}) &=& -4.88 \mbox{ deg,}
\\
\delta_{^3 P_1}  (25\mbox{ MeV}) &=& 2.56 \mbox{ deg.}
\end{eqnarray}
Finally, the contact parameter of the $^3 S_1- ^3 D_1$ transition potential,
$C_{^3 S_1- ^3 D_1}$ [Eq.~(\ref{eq_ct2_pw})], is adjusted such that 
the $\epsilon_1$ mixing parameter at 25 MeV reproduces the Nijmegen value,
\begin{equation}
\epsilon_1  (25\mbox{ MeV}) = 1.79 \mbox{ deg.}
\end{equation}

In previous studies, investigators looked at the
$NN$ phase shifts to judge if there was independence from the cutoff parameter
within the presumed accuracy~\cite{EGM00,NTK05,YEP09,EM06}. 
However, from phase shifts it is difficult to extract
the over-all error, because the deviations from the empirical phase
shifts can be very different in different partial waves, see Fig~\ref{fig_ph}.  
Moreover, it is well-known from the early
days of phase-shift analysis that there may exist several different phase-shift solutions
all of which fit the $NN$ data about equally well. 
This is due to the fact that the deviation from a particular phase shift solution
in one partial wave  may be compensated
by a deviation in some other partial wave. The only uniquely defined $NN$ data are
the original experimental data of $NN$ observable measurements.
Therefore, in the present investigation, we determine the accuracy of the predictions
by calculating the $\chi^2$ for the reproduction of the $NN$ data:
\begin{equation}
\chi^2 = \sum_{i=1}^{N} \left( \frac{x_i^{\rm theo} - x_i^{\rm exp}}{\Delta x_i^{\rm exp}} \right)^2  \,,
\label{eq_chi}
\end{equation}
where $N$ denotes the total number of experimental data (including normalizations),
$x_i^{\rm theo}$ is the prediction for datum $i$, and $x_i^{\rm exp}$ the experimental value
with uncertainty $\Delta x_i^{\rm exp}$. In general, we will state results for the $\chi^2$
in terms of $\chi^2/{\rm datum} \equiv \chi^2/N$.

To be more specific, we consider $np$ scattering and thus compare to the
$np$ data. The $np$ system has the advantage that it includes $I=0$ and $I=1$
and, thus, covers all $NN$ partial waves. We consider energies up to
pion-production theshold. Since the accuracy of the predictions by chiral EFT
is energy dependent, we subdivide the total energy range below pion-production
threshold into four
intervals and calculate the $\chi^2$ for those intervals.
The intervals are shown in the $\chi^2$ tables below.

\subsection{Renormalization at NLO}

\begin{table}
\caption{$\chi^2$/datum for the reproduction of the $np$ data at NLO 
for various values for $\Lambda$  of the regulator function
Eq.~(\ref{eq_f}).
Results are given for two choices for the SFR parameter, namely, $\tilde{\Lambda} = 700$ MeV and
 $\tilde{\Lambda} \rightarrow \infty$ (in parentheses).
$T_{\rm lab}$ denotes the kinetic energy of the incident neutron
in the laboratory system.
\label{tab_chiNLO}}
\smallskip
\begin{tabular}{ccccccc}
\hline \hline 
\noalign{\smallskip}
  & \multicolumn{6}{c}{$\Lambda$ (MeV)} \\
       \cline{2-7}
\noalign{\smallskip}
 $T_{\rm lab}$ bin (MeV)
 & 400
 & 500
 & 600
 & 700
 & 800
 & 900
\\
\hline \hline 
\noalign{\smallskip}
2--35&1.12&1.03&1.03&1.21&2.11&3.33\\
         &(1.17)&(1.04)&(1.07)&(1.25)&(1.94)&(2.85)\\
\hline 
\noalign{\smallskip}
 35--125&11.1&6.12&4.79&3.98&4.77&6.93\\
               &(12.0)&(6.44)&(4.98)&(4.16)&(5.63)&(8.64)\\
\hline 
\noalign{\smallskip}
 125--183&69.7&74.3&86.2&103&102&108\\
                &(75.9)&(76.5)&(88.3)&(95.3)&(96.8)&(104)\\
\hline \hline 
\end{tabular}
\end{table}

\begin{figure}
\vspace*{-3.5cm}
\hspace*{-1.0cm}
\includegraphics[scale=.5]{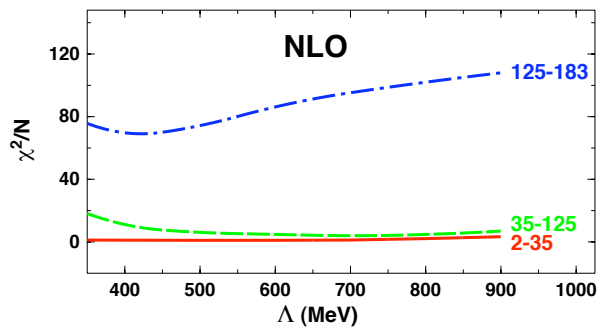}
\vspace*{-6.0cm}
\caption{(color online). $\chi^2$/datum for the reproduction of the $np$ data at NLO
as a function of the cutoff parameter $\Lambda$ of the regulator 
Eq.~(\ref{eq_f}). Results are shown for various
energy intervals as denoted in units of MeV.
The SFR parameter is always $\tilde{\Lambda}=700$ MeV.}
\label{fig_2}       
\end{figure}

\begin{table*}
\caption{Same as Table~\ref{tab_chiNLO}, but for NNLO.
The two choices for the SFR parameter are $\tilde{\Lambda} = 700$ MeV and
 $\tilde{\Lambda} = 600$ MeV (in parentheses).
\label{tab_chiNNLO}}
\smallskip
\begin{tabular*}{\textwidth}{@{\extracolsep{\fill}}ccccccccc}
\hline \hline 
\noalign{\smallskip}
  & \multicolumn{8}{c}{$\Lambda$ (MeV)} \\
       \cline{2-9}
\noalign{\smallskip}
 $T_{\rm lab}$ bin (MeV)
 & 400
 & 450
 & 500
 & 600
 & 700
 & 800
 & 850
 & 900
\\
\hline \hline 
\noalign{\smallskip}
2--35&0.91&0.88&0.87&0.87&0.87&0.87&0.87&0.98\\
         &(0.91)&(0.90)&(0.89)&(0.88)&(0.88)&(0.87)&(0.87)&(0.99)\\
\hline 
 35--125&6.36&2.80&1.96&1.88&2.27&2.24&2.23&2.21\\
               &(9.50)&(4.53)&(2.92)&(2.19)&(2.21)&(2.23)&(2.26)&(2.30)\\
\hline 
 125--183&52.2&15.0&6.40&4.33&6.22&8.75&9.73&16.4\\
                &(78.1)&(31.0)&(19.5)&(16.6)&(17.2)&(18.3)&(18.7)&(18.7)\\
\hline 
 183--290&149&40.9&18.9&18.0&23.1&27.8&28.2&50.2\\
                 &(194)&(76.2)&(52.1)&(54.8)&(56.1)&(54.4)&(52.0)&(48.4)\\
\hline \hline 
\end{tabular*}
\end{table*}

At NLO, we use $n=2$ in the regulator function Eq.~(\ref{eq_f}) and vary $\Lambda$
from 350 to 900 MeV. 
For reasons of efficiency, we calculate the $\chi^2$ at this order with the help
of the Nijmegen error matrix~\cite{SS93}. We have compared this approximate method
with an exact $\chi^2$ calculation and found that the error is only about $\pm 3$\%.
The results for the $\chi^2/{\rm datum} $ are shown in Table~\ref{tab_chiNLO} and Fig.~\ref{fig_2}. 
To get an idea for the 
dependence of our results on the SFR parameter $\tilde{\Lambda}$,
we conduct one series of $\Lambda$ variations with $\tilde{\Lambda}$ fixed at 700 MeV
and another one for $\tilde{\Lambda} \rightarrow \infty$, which is equivalent to dimensional regularization (DR). 
As mentioned, NLO does not really need SFR, which is why we include the case of DR.
As can be clearly seen from Table~\ref{tab_chiNLO}, the dependence on $\tilde{\Lambda}$
is very weak, which is comforting.

For the energy interval 2-35 MeV, the reproduction of the $np$ data is very good  for all cutoffs in the
range $\Lambda=350-800$ MeV where  $\chi^2/{\rm datum} \lea 2$ is obtained.
However, this should not come as a surprise since the $S$-waves are fit around zero energy
and the $P$-waves at 25 MeV. 
Thus, the results for the 2-35 MeV interval are not predictions.
The first predictions of the theory appear at the
interval 35-125 MeV. For $\Lambda$ between 500 to 850 MeV, the $\chi^2/{\rm datum}$
stays around $5 \pm 1$ which may be perceived as moderate cutoff independence.
Note that a $\chi^2/{\rm datum}$ around 5 signifies a good qualitative reproduction of the data.
Finally, in the third energy interval shown in Table~\ref{tab_chiNLO}, namely, 125-183 MeV,
the $\chi^2/{\rm datum}$ is 70 or larger, which simply means that the data are not reproduced.
Under these circumstances a discussion of $\Lambda$ independence is out of place.

In summary, chiral EFT at NLO is able to describe $NN$ scattering up to about 
100 MeV laboratory energy with moderate regulator independence for $\Lambda$'s
between 500 and 850 MeV.

\subsection{Renormalization at NNLO}

\begin{figure}
\vspace*{-1.5cm}
\hspace*{-1.0cm}
\includegraphics[scale=.5]{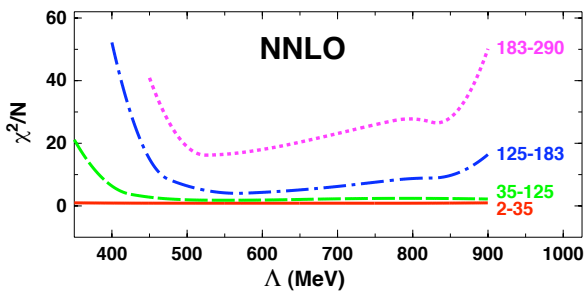}
\vspace*{-8.0cm}
\caption{(color online).  Same as Fig.~\ref{fig_2}, but for NNLO.}
\label{fig_3}       
\end{figure}

At NNLO, we use $n=3$ in the regulator function Eq.~(\ref{eq_f}) and 
calculate the $\chi^2$ exactly by applying Eq.~(\ref{eq_chi}) explicitly
to the 1999 data base~\cite{Mac01}, which includes  559 $np$ data
for the interval 2-35 MeV,  579 $np$ data for 35-125 MeV,  414 $np$ data for 125-183 MeV,
and  846 $np$ data for 183-290 MeV.
The results of these $\chi^2/{\rm datum}$ calculations are shown in Table~\ref{tab_chiNNLO}.
We also check the dependence of our results on the SFR parameter $\tilde{\Lambda}$,
for which we apply two choices, namely, 700 MeV and 600 MeV. The results for
$\tilde{\Lambda}=600$ MeV are given in parentheses.
The dependence on $\tilde{\Lambda}$ is moderate up to 125 MeV. However, significant differences
occur above 125 MeV with the choice $\tilde{\Lambda}=600$~MeV producing larger $\chi^2$.
This may be seen as an indication that $\tilde{\Lambda}=600$~MeV cuts out too much
from the 2PE contribution. We have also checked $\tilde{\Lambda}=800$~MeV, but found 
that spurious bound states and resonances start to occur, which SFR is supposed to prevent.
Thus, $\tilde{\Lambda}=700$~MeV appears to be the optimal choice at NNLO. 
This is not unreasonable on general grounds.
If the 2PE is described in terms of $2\pi$ resonances,  we have the $\sigma(600)$ with a mass
of 400-1200 MeV and the $\rho(770)$ with mass 775 MeV and width 150 MeV~\cite{Nak10}.
The choice $\tilde{\Lambda}=600$~MeV cuts out the contributions from all mass components
$\geq 600$~MeV, which obviously will trim the $\sigma$- and $\rho$-like contributions
severely. On the other hand, $\tilde{\Lambda}=700$~MeV will keep substantial contributions from $\sigma$
and $\rho$ alive.

The $\chi^2$ as a function of $\Lambda$ with $\tilde{\Lambda}$ fixed at 700~MeV
are plotted in Fig.~\ref{fig_3}.
The curves in this figure clearly reveal that, for all energy intervals above 35 MeV,
a cutoff $\Lambda \lea 450$~MeV cuts out too much from the intermediate range
part of the $NN$ potential, such that, particularly, the higher energies are not described well
(large $\chi^2$).
Then, on the other other end of the $\Lambda$ spectrum, the $\chi^2$ rise again
for $\Lambda \gea 850$~MeV, when too much of the unknown short-distance dynamics is
admitted.
In between the two extremes, namely for
$450 \; {\rm MeV} \lea \Lambda \lea 850 \; {\rm MeV}$,
one can clearly identify plateaus for all energy intervals.
This demonstrates cutoff independence for the physically relevant range and,
thus, successful renormalization.

At NNLO, the reproduction of the $NN$ data is acceptable up to about 200 MeV laboratory
energy, which is about 100 MeV above the limitations of the theory at NLO.

\subsection{Order by order improvement}

\begin{figure}
\vspace*{-2.0cm}
\hspace*{-0.5cm}
\includegraphics[scale=.5]{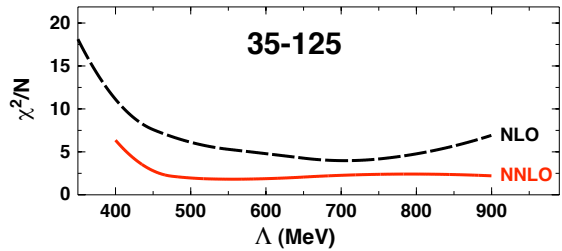}
\vspace*{-8.0cm}
\caption{(color online). $\chi^2$/datum for the reproduction of the $np$ data in the
energy range 35-125 MeV
as a function of the cutoff parameter $\Lambda$ of the regulator function
Eq.~(\ref{eq_f}). The (black) dashed curve shows the $\chi^2$/datum
achieved with a potential constructed at order NLO and the (red) solid curve is for NNLO.
The SFR parameter is always $\tilde{\Lambda}=700$ MeV.}
\label{fig_32}       
\end{figure}

\begin{figure}
\vspace*{-1.0cm}
\hspace*{-0.5cm}
\includegraphics[scale=.5]{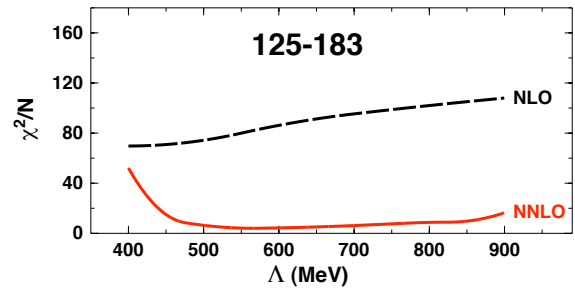}
\vspace*{-8.0cm}
\caption{(color online). Same as Fig.~\ref{fig_32}, but for the
energy range 125-183 MeV.}
\label{fig_32b}       
\end{figure}

Besides regulator independence, a proper EFT should also show
order-by-order improvement of the predictions with decreasing error.
We address this issue in Fig.~\ref{fig_32} for the energy interval 35-125 MeV
and in Fig.~\ref{fig_32b} for 125-183 MeV. In each figure, we show the NLO and NNLO 
$\chi^2$ results for easy comparison. The figures clearly demonstrate that, 
when going from NLO to NNLO, the $\chi^2$ is drastically reduced and,
at the same time, the $\Lambda$-independence substantially improved.
This is exactly what one wants to see in an EFT.

It is important to stress that NLO and NNLO have the same number of contact
parameters. Thus, the improvements seen at NNLO 
are not due to a larger number of fit parameters. 
The difference at NNLO is an improved 2PE contribution to the
$NN$ interaction. At this order, the subleading $\pi\pi NN$ vertices from the dimension-two
Lagrangian with LECs $c_i$  contribute [cf.\ Eqs.~(\ref{eq_3C})-(\ref{eq_3T})].
These vertices represent correlated 2PE
as well as intermediate $\Delta(1232)$-isobar excitation.
It is well-known from the conventional meson theory of 
nuclear forces~\cite{MHE87}
that these two mechanisms are crucial
for a realistic and quantitative 2PE model.
Consequently, at NNLO, the chiral 2PE  assumes a realistic size
and describes the intermediate-range attraction of the
nuclear force right. 
This explains the improvements seen at NNLO, which are
 demonstrated in a particularly impressive way in Fig.~\ref{fig_32b}.

\section{Conclusions}

We have investigated the nonperturbative renormalization of nucleon-nucleon scattering
at next-to-leading order (NLO) and next-to-next-to-leading order (NNLO) of chiral effective field theory.
We keep the cutoff parameter $\Lambda$ below the chiral symmetry breaking scale of about 1~GeV.
The accuracy of the fits is measured by the $\chi^2$ for the reproduction of the $NN$ data.
Applying this measure, we find that, at NLO, the $NN$ data are described well up to about
100 MeV laboratory energy and, at NNLO, up to about 200 MeV.
Concerning the cutoff dependence, we can clearly identify plateaus of insensitivity to 
changes of $\Lambda$ for 
$450 \; {\rm MeV} \lea \Lambda \lea 850 \; {\rm MeV}$.
We perceive this result as successful renormalization of the chiral $NN$ interaction
at the two orders considered in this study.

\section*{Acknowledgements}
This work was supported in part by the U.S. Department of Energy
under Grant No.~DE-FG02-03ER41270.
The work of D. R. E. was funded by the Ministerio de Ciencia y
Tecnolog\'\i a under Contract No.~FPA2010-21750-C02-02 and
the European Community-Research Infrastructure Integrating
Activity ``Study of Strongly Interacting Matter'' (HadronPhysics3
Grant No.~283286).


\begin{thebibliography}{99}
\bibitem{ME11}
R. Machleidt and D. R. Entem, Phys. Rep. {\bf 503}, 1 (2011).
\bibitem{EHM09}
E. Epelbaum, H.-W. Hammer, and U.-G. Mei\ss ner,
Rev. Mod. Phys. {\bf 81}, 1773 (2009).
\bibitem{GL84}
J. Gasser and H. Leutwyler,
Ann. Phys. (N.Y.) {\bf 158}, 142 (1984).
\bibitem{GSS88}
J. Gasser, M. E. Sainio, and A. \v{S}varc,
Nucl. Phys. {\bf B307}, 779 (1988).
\bibitem{Wei90} S. Weinberg, Phys. Lett {\bf B251}, 288 (1990); Nucl.\ Phys.\ {\bf B363}, 3 (1991).
\bibitem{ORK94}
C. Ord\'o\~nez, L. Ray, and U. van Kolck,
Phys.\ Rev.\ Lett.\ {\bf 72}, 1982 (1994);
Phys.\ Rev.\ C {\bf 53}, 2086 (1996).
\bibitem{KBW97} N. Kaiser, R. Brockmann, and W. Weise,
Nucl.\ Phys.\ {\bf A625}, 758 (1997).

\bibitem{EGM98} E. Epelbaum, W. Gl\"ockle, and U.-G. Mei\ss ner,
Nucl.\ Phys.\ {\bf A637}, 107 (1998).
\bibitem{EGM00} E. Epelbaum, W. Gl\"ockle, and U.-G. Mei\ss ner,
Nucl.\ Phys.\ {\bf A671}, 295 (2000).



\bibitem{EM02a} D. R. Entem and R. Machleidt,
Phys. Rev. C {\bf 66}, 014002 (2002).
\bibitem{EM03} D. R. Entem and R. Machleidt,
Phys. Rev. C {\bf 68}, 041001 (2003).
\bibitem{EGM05} E. Epelbaum, W. Gl\"ockle, and U.-G. Mei\ss ner,
Nucl. Phys. {\bf A747}, 362 (2005).

\bibitem{MHE87} 
R. Machleidt, K. Holinde, and Ch.\ Elster,
Phys.\ Rep.\ {\bf 149} (1987) 1.

\bibitem{Wei09}
S. Weinberg,
{\it Effective Field Theory, Past and Future}, 
arXiv:0908.1964 [hep-th].
\bibitem{KSW96}
D. B. Kaplan, M. J. Savage, and M. B. Wise,
Nucl. Phys. {\bf B478}, 629 (1996).
\bibitem{NTK05}
A. Nogga, R. G. E. Timmermans, and U. van Kolck, 
Phys. Rev. C {\bf 72}, 054006 (2005).
\bibitem{YEP09}
C.-J. Yang, Ch. Elster, D. R. Phillips, 
Phys. Rev. C {\bf 80}, 044002 (2009).
\bibitem{VA07}
M. Pavon Valderrama, and E. Ruiz Arriola,
Phys. Rev. C {\bf 74}, 064004 (2006).
\bibitem{Ent08} 
D. R. Entem, E. Ruiz Arriola, M. Pav\'on Valderrama, and R. Machleidt, 
Phys. Rev. C {\bf 77}, 044006 (2008)
\bibitem{ZME12}
Ch. Zeoli, R. Machleidt, D. R.  Entem, 
Few-Body Syst. DOI 10.1007/s00601-012-0481-4, arXiv:1208.2657 [nucl-th].
\bibitem{EG09}
E. Epelbaum and J. Gegelia, 
Eur. Phys. J. {\bf A41}, 341 (2009).
\bibitem{EG12}
E. Epelbaum and J. Gegelia, 
Phys. Lett B {\bf 716}, 338 (2012).
\bibitem{LK08} 
B. Long, and U. van Kolck, 
Ann. Phys. (N.Y) {\bf 323}, 1304 (2008)
\bibitem{Val11}
M. P. Valderrama,  
Phys. Rev. C {\bf 84}, 064002 (2011)
\bibitem{Mac09}
R. Machleidt, P. Liu, D. R.  Entem, and E. R. Arriola, 
Phys. Rev. C {\bf 81}, 024001 (2010)
\bibitem{Lep97}
G. P. Lepage, 
{\it How to Renormalize the Schr\"odinger Equation},
arXiv:nucl-th/9706029.
\bibitem{EGM04}
E. Epelbaum, W. Gl\" ockle, and U.-G. Mei\ss ner,
Eur. Phys. J. {\bf A19}, 401 (2004).
\bibitem{BM00} 
P. B\"{u}ttiker and U.-G. Mei\ss ner,
Nucl.\ Phys.\ {\bf A668}, 97 (2000).
\bibitem{BS66} 
R. Blankenbecler and R. Sugar, 
Phys.\ Rev.\ {\bf 142}, 1051 (1966).
\bibitem{EM06}
E. Epelbaum and U.-G. Mei\ss ner,
{\it On the renormalization of the one-pion exchange potential and the consistency of
Weinberg's power counting},
arXiv:nucl-th/0609037.

\bibitem{Sto93} V.\ G.\ J.\ Stoks, R.\ A.\ M.\ Klomp, 
M.\ C.\ M.\ Rentmeester, and J.\ J.\ de Swart, 
Phys.\ Rev.\ C {\bf 48}, 792 (1993).
\bibitem{SM99} 
R. A. Arndt, I. I. Strakovsky, and R. L. Workman,
SAID, Scattering Analysis Interactive Dial-in computer facility,
George Washington University
(formerly Virginia Polytechnic Institute), 
solution SM99 (Summer 1999); for more information see, e.~g.,
R. A. Arndt, I. I. Strakovsky, and R. L. Workman,
Phys. Rev. C {\bf 50}, 2731 (1994).

\bibitem{SS93}
V. Stoks and J. J. de Swart,
Phys. Rev. C {\bf 47}, 761 (1993); 
and V. Stoks, private communication.

\bibitem{Mac01} 
The 1999 data base is defined in:
R. Machleidt, Phys. Rev. C {\bf 63} 024001 (2001).

\bibitem{Nak10}
K. Nakamura {\it et al.}, J. Phys. G: Nucl. Part. Phys. {\bf 37}, 075021 (2010).

\end{thebibliography}
\end{document}